\documentclass[10pt,conference]{IEEEtran}

\usepackage{cite}
\usepackage{amsmath,amssymb}
\usepackage{booktabs}
\usepackage{array}
\usepackage[T1]{fontenc}
\usepackage[utf8]{inputenc}
\usepackage{url}
\usepackage{xcolor}
\usepackage{tikz}
\usetikzlibrary{positioning,arrows.meta}
\usepackage{pgfplots}
\pgfplotsset{compat=1.18}
\definecolor{qcBlue}{HTML}{2A78D6}
\definecolor{qcOrange}{HTML}{EB6834}
\definecolor{qcGreen}{HTML}{1B9E77}
\usepackage[colorlinks=true,allcolors=blue]{hyperref}

\newcommand{\id}[1]{\texttt{#1}}
\providecommand{\tightlist}{\setlength{\itemsep}{0pt}\setlength{\parskip}{0pt}}
\makeatletter\renewcommand\verbatim@font{\scriptsize\ttfamily}\makeatother

\begin{document}

\title{Design-Time Conformance Checking for\\Pulse-Level Quantum Control}

\author{\IEEEauthorblockN{Rylan Malarchick}
\IEEEauthorblockA{\textit{Department of Engineering Physics} \\
\textit{Embry-Riddle Aeronautical University}\\
Daytona Beach, FL, USA \\
\href{mailto:malarchr@my.erau.edu}{malarchr@my.erau.edu}}
}

\maketitle

\begin{abstract}
A pulse-level quantum control program is written against a device whose limits are recorded, if at all, in vendor documentation and source code. A program that exceeds them can be refused at compile time. It can also be accepted and silently altered, or compile and then fail at the board. In the last two cases the experiment runs, and the data does not correspond to the program that was written.

We present qconform, a checker that decides whether a pulse program is realizable on a device, given a versioned capability descriptor for that device. Every constraint in a descriptor cites the toolchain observation that established it. The checker is design-time, offline, and deterministic, and it uses no floating point. It reports a verdict, the rules it applied, and a coverage manifest that names what it did not check.

We evaluate qconform by differential testing against the QICK and Qblox toolchains, on three QICK board configurations and one Qblox cluster. Over 1263 program runs of 969 distinct programs, the checker accepted no program that a toolchain refuses. On both vendors, a descriptor pinned to one toolchain release accepts programs that older releases refuse. The evaluation also found ten defects in the checker and its descriptors, all fixed.

\end{abstract}

\begin{IEEEkeywords}
quantum control, pulse-level programming, conformance checking, differential
testing, capability description
\end{IEEEkeywords}

\hypertarget{sec:intro}{%
\section{Introduction}\label{sec:intro}}

Pulse-level control stacks expose a programming interface whose constraints are
not part of the interface. A generator has a minimum pulse length, a timing
grid, a frequency band, an amplitude resolution, and a finite envelope memory.
None of these appear in the function signature that the user calls. They appear
as an exception at compile time, as a silently quantized value, or as a failure
at board load.

The silent cases matter most. If a toolchain accepts a 60.18-cycle pulse and
runs 60 cycles, the experiment produces data. Nothing in the data says that the
pulse was shortened.

We treat the limits of a device as a contract. A constraint written in prose
drifts from the toolchain that enforces it. A constraint in a versioned,
machine-readable descriptor cites an observation, can be regenerated when the
toolchain changes, and can be checked mechanically. A checker then decides each
program against the descriptor before the program runs.

This paper presents qconform, a checker built on that contract. It contributes:

\begin{enumerate}
\def\labelenumi{\arabic{enumi}.}
\tightlist
\item
  A capability descriptor format in which every constraint cites the survey
  observation that established it. A gate refuses a descriptor whose citations
  do not resolve (Section~\ref{sec:descriptor}).
\item
  A checker in C99 with no dependencies. It decides a program against a
  descriptor and reports what it checked as well as what it found (Section~\ref{sec:checker}).
\item
  A differential evaluation against the QICK and Qblox toolchains on four
  configurations. The checker accepted no program that a toolchain refuses,
  over 1263 program runs. The toolchain version pin changes that result on
  both vendors (Section~\ref{sec:eval}).
\end{enumerate}

\hypertarget{sec:problem}{%
\section{What a toolchain does silently}\label{sec:problem}}

We surveyed the QICK asm\_v2 toolchain by black-box probing. The survey drives the
toolchain from captured board configurations, with no hardware present. Each
probe that runs records one of four outcomes. The toolchain accepts the program unchanged,
accepts it and changes a value, refuses it with a typed error, or fails without
naming a cause. A user sees the third and fourth outcomes. The second is
invisible.

The survey has 605 probes over three configurations, written as 45 catalog
files. Each probe varies one parameter around a limit that the configuration
implies. The catalog therefore records where each limit sits, and on which side
the toolchain refuses, repairs, or stays silent.

Three QICK behaviors shape the rest of the paper. On a full-speed generator, a
frequency past the range of the digital synthesizer is accepted and aliased, and
the register wraps. A gain past full scale is accepted, and the register exceeds what the hardware can
represent. An envelope larger than channel memory compiles, and the failure
appears at board load. In each case the toolchain accepts the program and the
hardware cannot honor it.

The tone table of a muxed generator shows how deep a silent repair can sit. A
tone one step below the lower band edge compiles and is folded onto its Nyquist
image, about 860 MHz from the request. The readback reports the requested
frequency. Only the register shows the repair. The toolchain also rounds
readout lengths silently, to whole readout cycles, half to even.

We surveyed the Qblox toolchain the same way. qblox-scheduler compiles a
schedule to Q1ASM for a dummy cluster that qblox-instruments runs with no
hardware. That survey has 193 probes over one configuration, in 11 catalog
files. It found the same pattern. A clock declared with an initial phase
compiles, and the sequencer starts at phase zero with no warning. An amplitude
of exactly full scale is written as 32767 of 32768, because the register cannot
hold +1. An amplitude below half a register step rounds to zero, and the play
instruction disappears from the program. Operations on one sequencer must start
at the same time or at least 4 ns apart.

\hypertarget{sec:descriptor}{%
\section{The capability descriptor}\label{sec:descriptor}}

A descriptor names a device and lists its channels. Each channel declares a time
base, two grids, its capabilities, and its constraints. The descriptor is pinned
to a firmware configuration and a library version. The survey showed that
accept and reject behavior is a function of that pair.

The time base is an exact rational number of seconds. It is chosen so that one
channel clock cycle and one processor timing tick are both whole multiples of
it. The two clocks differ. On the QICK board we study, a generator runs a
599.04 MHz fabric clock against a 430.08 MHz timing clock. A single grid field
would misstate the device.

Every constraint carries an evidence list that points into the survey catalog.
A gate script refuses a descriptor whose evidence does not resolve. The
frequency constraint of an interpolated generator, for example, reads:

\begin{verbatim}
{"id": "frequency_range",
 "shape": "range_resolution",
 "severity": "fatal", "post_mixer": true,
 "min": {"num": -860160000, "den": 1},
 "max": {"num": 28185722866875,
         "den": 32768},
 "resolution": {"num": 13125, "den": 32768},
 "evidence": [
  {"axis": "freq", "outcome": "reject",
   "note_contains": "just under lower edge"},
  {"axis": "freq", "outcome": "accept",
   "note_contains": "lower edge"},
  {"axis": "freq", "outcome": "reject",
   "note_contains": "upper edge"},
  {"axis": "freq", "outcome": "reject",
   "note_contains": "1.5x f_dds"}]}
\end{verbatim}

Values are exact rationals in SI units. The checker compares the band after it
subtracts the mixer. The minimum is the lower band edge. The maximum is one
resolution step inside the upper edge. The toolchain subtracts the mixer in
double precision, and a request at exactly the upper edge becomes
860.1600000000001 MHz, which it refuses. Each evidence entry names the survey
rows that show one of these facts. The configuration field of each entry is
omitted here.

The second device is a Qblox cluster, one QCM module for drive and one QRM for
readout. Its architecture differs from QICK, and it uses parts of the format
that the QICK descriptors do not.

A frame is one port and one clock, which the toolchain compiles to one
sequencer. The spacing between operations, the waveform memory, and the
instruction budget are therefore scoped to a frame. Two frames on one module
each carry their own count.

The channel unit is one nanosecond and both grids are one unit. The 4 ns figure
in the vendor documentation is a minimum spacing between operation starts. An
envelope of 5, 6, 7, 9, or 12
samples compiles and reads back its own sample count, so the sample grid is one
sample.

Six constraint forms come from this device. \texttt{start\_spacing},
\texttt{frequency\_update\_spacing}, and \texttt{capture\_spacing} bound the time between
operations of one kind on a frame. \texttt{capture\_slot} states that a capture
occupies a fixed slot whatever it integrates. \texttt{capture\_length\_uniform} states
that one sequencer holds one integration length. \texttt{envelope\_duration\_exact}
states that a play of a waveform lasts exactly its sample count, because this
device neither pads nor truncates. A constraint may also declare saturation.
The toolchain clamps an amplitude between 32767/32768 of full scale and full
scale, so the checker reports it as a repair.

\hypertarget{sec:checker}{%
\section{The checker}\label{sec:checker}}

The checker reads a descriptor and a program, both JSON, and emits a JSON
report. It has four exit codes: pass, fail, pass with repairs, and tool error.

\textbf{Arithmetic is exact.} Durations are integer counts of a rational time base,
and frequencies, phases, and amplitudes are rationals. Neither the program
format nor the checker uses floating point, so a grid test is exact.

\textbf{A verdict and a coverage manifest are reported together.} Every report
classifies all 26 coverage classes as checked, not applicable, unchecked, or
indeterminate, so a pass states its own scope. A channel that a descriptor
declares but does not constrain reports as unchecked, which separates a vacuous
pass from a verified one.

\textbf{A repair is a verdict of its own.} Each rejection is fatal or
vendor-repairable. A repairable rejection names what the toolchain will change,
and the verdict is pass with repairs. A user sees the silent case before the
program runs.

The checker is 4488 lines of C99 with no dependencies. Building and testing it
require a C compiler and a POSIX shell.

{\raggedright\emph{Source: \texttt{src/}, \texttt{documentation/spec-v0.txt}.}\par}

\hypertarget{sec:scope}{%
\section{What qconform does not do}\label{sec:scope}}

The checker makes one claim: if it returns pass or pass with repairs, the
pinned vendor toolchain compiles the program. The items below are outside that
claim.

\textbf{What the checker does not decide.}

\begin{itemize}
\tightlist
\item
  \textbf{The converse.} The checker may refuse a program that the toolchain
  compiles. Three QICK cases are deliberate, because the toolchain accepts what
  the hardware cannot honor: an aliased frequency, a gain past full scale, and
  an envelope past channel memory. In a fourth case, on both vendors, the checker refuses a frequency or phase update
  that no later output uses. The toolchain checks only what an output uses.
\item
  \textbf{Another toolchain release.} Section~\ref{sec:versions} shows that a verdict does not
  transfer.
\item
  \textbf{Run-time behavior.} The checker runs before the program does. The program
  format has no branches and no loops, so data-dependent control flow is out of
  scope.
\item
  \textbf{Accumulated time on a muxed QICK generator.} The toolchain bounds
  accumulated program time. The rule bounds a single pulse. It is exact for a
  pulse that starts the program and loose for one that starts later.
\item
  \textbf{Other input languages.} Programs arrive in the qconform JSON format. One
  exporter exists, from compiled QICK programs.
\end{itemize}

\textbf{What the evaluation does not show.}

\begin{itemize}
\tightlist
\item
  \textbf{Absence.} A clean corpus does not prove that no unsound program exists.
\item
  \textbf{Static QICK readouts.} Such a readout is configured outside the program,
  and the harness does not lower a capture on one. This affects one row of 681.
\item
  \textbf{Two Qblox grid rules.} The channel unit is one nanosecond and both grids
  are one unit, so only a program stated in half a unit can miss either grid.
  The vendor interface takes seconds as a double, and a half-nanosecond value
  does not survive that conversion exactly, so neither rule is tested against
  the toolchain.
\item
  \textbf{Limits the toolchain does not check.} On QICK, the mux tone count and the
  envelope memory are enforced from the descriptor alone, with no oracle.
\item
  \textbf{Other vendors.} Both vendors here are control stacks for superconducting
  qubits, and both compile with no hardware present. A vendor whose compiler
  runs only against a live instrument cannot be tested this way.
\end{itemize}

\hypertarget{sec:eval}{%
\section{Differential evaluation}\label{sec:eval}}

\hypertarget{sec:method}{%
\subsection{Method}\label{sec:method}}

The oracle is the vendor toolchain itself, run with no hardware. QICK compiles
against a board configuration captured from a published board. Qblox compiles
for a dummy cluster built from a hand-written hardware configuration. A QICK
compile takes about 1.9 ms, and a Qblox compile a median of 11 ms. Corpus size is
therefore not a constraint.

The harness lowers each generated program into a vendor program and compiles it.
It records the verdict and the vendor outcome together, one row per program.

Both vendor interfaces take times and frequencies as doubles, and the checker is
exact. Every conversion records the request, the value passed, and whether the
two fall in the same grid cell. Triage attributes a row where they do not to the
harness, and excludes it from the soundness count.

The corpus has two parts. Boundary ladders walk each declared limit at far
below, just below, exactly at, just above, and far above. Budget ladders probe
exactly the limit and one element past it. Spacing ladders place two operations
one unit either side of the minimum spacing. Resolution and grid rungs include an
odd multiple of the step, which is on the declared lattice and off a lattice
twice as coarse. Randomized programs compose several elements with values drawn
near the limits. They find combinations that the ladders do not imply.

The corpus probes one representative of each channel class. Two channels are in
one class only when every value the descriptor gives them is the same. On the
largest QICK configuration this gives eight classes among 27 channels. On Qblox
each of the three channels is its own class.

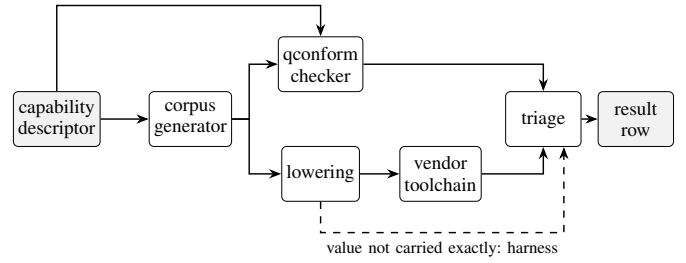
\begin{figure}[t]
  \centering
  \resizebox{\columnwidth}{!}{\begin{tikzpicture}[
    box/.style={draw, rounded corners=2pt, align=center, font=\footnotesize,
                minimum height=8mm, minimum width=11mm, inner sep=2pt},
    data/.style={box, fill=black!5},
    arr/.style={-{Stealth[length=1.8mm]}, line width=0.6pt},
]
\node[data] (desc) at (0,0)    {capability\\descriptor};
\node[box]  (gen)  at (1.95,0)  {corpus\\generator};
\node[box]  (chk)  at (3.85,0.8) {qconform\\checker};
\node[box]  (low)  at (3.85,-0.8) {lowering};
\node[box]  (orc)  at (5.6,-0.8) {vendor\\toolchain};
\node[box]  (tri)  at (7.1,0)  {triage};
\node[data] (row)  at (8.45,0) {result\\row};
\draw[arr] (desc) -- (gen);
\draw[arr] (gen.east) -- ++(0.25,0) |- (chk.west);
\draw[arr] (gen.east) -- ++(0.25,0) |- (low.west);
\draw[arr] (desc.north) -- ++(0,1.25) -| (chk.north);
\draw[arr] (low) -- (orc);
\draw[arr] (chk.east) -| (tri.north);
\draw[arr] (orc.east) -| (tri.south);
\draw[arr] (tri) -- (row);
\draw[arr, dashed] (low.south) -- ++(0,-0.45) -| node[pos=0.25, below, font=\scriptsize]
    {value not carried exactly: harness} ([xshift=3mm]tri.south);
\end{tikzpicture}}
  \caption{The differential loop. The generator derives programs from the
  descriptor, the checker and the vendor toolchain each answer, and triage
  dispositions every row. A value the lowering cannot carry exactly is
  attributed to the harness, not counted as agreement or disagreement.}
  \label{fig:loop}
\end{figure}

\hypertarget{sec:dispositions}{%
\subsection{Dispositions}\label{sec:dispositions}}

Only an unsound pass, a program the checker accepts and the toolchain refuses,
counts against the claim.

Triage puts each row in one of nine classes. It compares repairs one quantity
at a time (length, start time, frequency, phase, gain).

\begin{itemize}
\tightlist
\item
  \textbf{agree}: both answers say the same thing, quantity by quantity.
\item
  \textbf{unsound}: the checker accepted and the vendor refused.
\item
  \textbf{missed repair}: the vendor changed a quantity that the checker predicted
  no repair to.
\item
  \textbf{over-predicted}: the checker predicted a repair to a quantity that the
  vendor left unchanged.
\item
  \textbf{vendor lenient}: the checker flagged what the vendor compiles, and the
  descriptor documents the later failure on that channel.
\item
  \textbf{conservative}: the checker refused and the vendor compiled, and every
  refusal points at a frequency or phase update that no later output uses.
\item
  \textbf{unobserved}: a repair that the harness cannot see.
\item
  \textbf{harness}: a row that the harness could not express, or a value that it
  could not carry exactly.
\item
  \textbf{open}: anything else.
\end{itemize}

\hypertarget{sec:results}{%
\subsection{Results}\label{sec:results}}

\begin{table*}[t]
  \centering
  \caption{Differential result on four configurations, three QICK and one Qblox (qick 0.2.418, qblox-scheduler \textnormal{1.0.0b8}, seed 1).}
  \label{tab:results}
  \begin{tabular}{@{}lrrrrrrr@{}}
  \toprule
  Configuration & Programs & Unsound & Agree & Vendor lenient & Conservative & Harness & Coverage \\
  \midrule
  testbench & 127 & 0 & 107 & 15 & 0 & 5 & 15/15 \\
  qce2025-r26 & 681 & 0 & 598 & 61 & 2 & 20 & 17/17 \\
  rb-r27 & 127 & 0 & 107 & 15 & 0 & 5 & 15/15 \\
  qblox-qcm-qrm & 328 & 0 & 312 & 0 & 6 & 10 & 19/19 \\
  \bottomrule
  \end{tabular}
\end{table*}

No configuration has an unsound pass, a missed or over-predicted repair, an
unobserved repair, or an open row.

rb-r27 is the testbench board with four times the program memory. It shares its
channel structure and all but four of its distinct programs with testbench.
Across the four configurations there are 1263 runs of 969 distinct programs:
665 on QICK and 304 on Qblox. The linear program-memory cost model holds
exactly at both QICK memory sizes. The program one element past the limit
compiles to 4097 of 4096 words on testbench, and to 16385 of 16384 words on
rb-r27. On Qblox the instruction cost model holds on both budgets, one per
module type. The program one element past the limit compiles to 16386 of 16384
instructions on the drive module. On the readout module it compiles to 12290 of
12288.

\begin{table*}[t]
  \centering
  \caption{Rule classes: programs where each fires and where it is checked and
  holds. rb-r27 matches testbench class for class and is not shown. A dash is a
  class the configuration's channels do not declare. ``vacuous'' marks a class
  the descriptor states in a form that can never reject.}
  \label{tab:coverage}
  \footnotesize
  \begin{tabular}{@{}lrrrrrr@{}}
\toprule
 & \multicolumn{4}{c}{QICK} & \multicolumn{2}{c}{Qblox} \\
\cmidrule(lr){2-5}\cmidrule(lr){6-7}
 & \multicolumn{2}{c}{testbench} & \multicolumn{2}{c}{qce2025-r26} & \multicolumn{2}{c}{qblox-qcm-qrm} \\
\cmidrule(lr){2-3}\cmidrule(lr){4-5}\cmidrule(lr){6-7}
rule class & fires & holds & fires & holds & fires & holds \\
\midrule
\id{amplitude\_range} & 26 & 91 & 138 & 493 & 57 & 249 \\
\id{amplitude\_resolution} & 5 & 112 & 230 & 401 & 2 & 304 \\
\id{capture\_length\_uniform} & -- & -- & -- & -- & 2 & 98 \\
\id{capture\_slot} & -- & -- & -- & -- & 10 & 90 \\
\id{capture\_spacing} & -- & -- & -- & -- & 4 & 96 \\
\id{envelope\_amplitude} & 2 & 8 & 10 & 40 & 6 & 16 \\
\id{envelope\_duration\_exact} & -- & -- & -- & -- & 2 & 20 \\
\id{envelope\_memory} & 1 & 9 & 5 & 45 & 2 & 20 \\
\id{envelope\_sample\_grid} & 2 & 8 & 6 & 44 & \multicolumn{2}{c}{vacuous} \\
\id{frequency\_range} & 6 & 121 & 67 & 614 & 35 & 293 \\
\id{frequency\_resolution} & 1 & 126 & 6 & 675 & 4 & 324 \\
\id{frequency\_update\_spacing} & -- & -- & -- & -- & 14 & 88 \\
\id{mux\_tone\_count} & -- & -- & 1 & 68 & -- & -- \\
\id{mux\_tone\_mask} & -- & -- & 2 & 67 & -- & -- \\
\id{negative\_duration} & 1 & 126 & 6 & 675 & 2 & 326 \\
\id{phase\_resolution} & 2 & 125 & 11 & 670 & 34 & 294 \\
\id{pmem\_words} & 2 & 124 & 2 & 678 & 4 & 320 \\
\id{pulse\_length\_grid} & 17 & 109 & 61 & 613 & 4 & 322 \\
\id{pulse\_length\_range} & 19 & 107 & 102 & 573 & 25 & 301 \\
\id{readout\_length\_range} & 4 & 14 & 20 & 70 & 8 & 92 \\
\id{schedule\_grid} & 50 & 77 & 142 & 539 & 4 & 324 \\
\id{start\_spacing} & -- & -- & -- & -- & 65 & 263 \\
\id{wmem\_words} & 2 & 125 & 2 & 679 & -- & -- \\
\bottomrule
\end{tabular}

\end{table*}

A class counts as covered when the corpus has a program in which it fires and
one in which it is checked and holds. A configuration reaches only the classes
that its channels declare. Of the 26 classes, 23 are exercised. Two of the other three are unreachable by construction, and one
reports unchecked by definition. On Qblox the envelope sample grid is one sample
on every channel, so that class can never reject. The table reports it as
vacuous.

The 91 vendor-lenient QICK rows are behaviors that the descriptor documents.
In 83 of them, an aliased or folded frequency or a gain past full scale, only
the raw register shows the change, because the readback parameter reports the
requested value. The other 8 exceed a limit that the toolchain does not check:
envelope memory or the mux tone count. A frequency just past the band reads
back unchanged, and its register is 2147483651, which has wrapped past \(2^{31}\).
No Qblox row is lenient. That toolchain clamps an amplitude of exactly full
scale, and the checker predicts the clamp as a saturation repair.

The eight conservative rows each set a frequency or a phase that no later output
uses.

Triage attributes 40 rows to the harness. One is a capture on a static QICK
readout. Two are Qblox programs whose play lasts one
sample longer than its waveform, which the Qblox lowering cannot express. The
other 37 are values that did not survive the double that the vendor interface
takes.

Only the randomized programs depend on the corpus seed. Over ten seeds the
corpus grows to 486, 2564, and 486 distinct programs on the three QICK
configurations, and to 1024 on Qblox. No seed gives an unsound pass, a missed
repair, an over-predicted repair, or an open row.

{\raggedright\emph{Source: \texttt{tools/differential/results/*.jsonl}, qick 0.2.418 and
qblox-scheduler 1.0.0b8, seed 1, reproduced with \texttt{make\ differential} and
\texttt{make\ differential-qblox}. Seeds: qconform-study \texttt{saturation/results/}.}\par}

\hypertarget{sec:versions}{%
\subsection{The toolchain version pin}\label{sec:versions}}

The descriptor pin is part of the soundness claim on both vendors. We reran the
survey and the differential under four other releases of each toolchain, and
held everything else fixed.

On QICK, two releases, one older and one newer than the pin, give the same result
rows byte for byte. The two oldest do not. The oldest refuses the newer
processor revision outright. The next refuses any jump target past address 2047,
so it refuses a program at the program-memory limit that the pinned descriptor
passes. Against the pinned descriptors, the oldest release gives 90, 2, and 90
unsound passes on the three configurations, and the next gives 2 on each.

On Qblox, the pin is the newest release, so all four others are older. The
nearest one reproduces the catalog and the results byte for byte. The three
oldest refuse a wait shorter than 4 ns inside a program, which the pin accepts.
Two of them also compile the program at each instruction limit to one
instruction over it. Against the pinned descriptor, they give 10, 10, and 8
unsound passes. The three oldest also need an older release of a support
library, so there the scheduler is not the only change.

{\raggedright\emph{Source: qconform-study \texttt{versions/results/}.}\par}

\hypertarget{sec:defects}{%
\subsection{Defects the evaluation found}\label{sec:defects}}

The differential and a fault-injection study of the harness found defects in
three places. Fault injection planted 742 single-change defects in the
descriptors and the checker. With the corpus built from the correct descriptor,
six on a probed channel survive for lack of a test program.

\begin{table*}[t]
  \centering
  \caption{Defects the evaluation found, by where they sat.}
  \label{tab:defects}
  \footnotesize
  \begin{tabular}{@{}llp{0.6\textwidth}@{}}
  \toprule
  Kind & Count & Example \\
  \midrule
  checker and descriptors & 10 (7 QICK, 3 Qblox) & The Qblox descriptor charged one instruction per output. The toolchain emits two, and the checker passed a program over the instruction limit. \\
  harness & 18 (15 QICK, 3 Qblox) & The lowering defined a new pulse for every play, which exhausted the vendor waveform table and looked like an unsound pass. \\
  evidence & 5 & The corpus seed came from a salted string hash. Of 286 programs, 120 differed between two runs at the same seed. \\
  \bottomrule
  \end{tabular}
\end{table*}

The differential found seven of the ten checker and descriptor defects, and
fault injection found three. Each is fixed.

One kind of defect recurred three times: a descriptor field that the checker
accepted and never enforced. The descriptor recorded that an interpolated QICK
generator checks its band after the digital mixer, as a \texttt{post\_mixer} flag. The
parser read the flag, and the checker never consulted it. It compared the
requested frequency against an unshifted band and accepted four programs that
the toolchain refuses. The fix was a format change. A program now declares the
mixer of its channel, and the checker refuses a post-mixer program that
declares none.

The second instance was the tone count of a muxed generator: declared,
schema-checked, parsed, and read by nothing. We then added a build-time
tripwire that requires the checker to read every descriptor field that the
parser fills. It fires when the mixer flag is removed from the checker. On its first run it found three more
fields in the same state, which are recorded as open.

The third instance was the Qblox instruction cost in the table above. The
evidence gate resolved each citation of the cost model. It never asked the
model to predict the cited rows, which show two instructions per output. The
gate now requires each linear cost model to predict the accept or refuse outcome of every
row that its evidence cites. With the old cost restored, it refuses the
descriptor and names both rows.

{\raggedright\emph{Source: commits \texttt{fd09d08}, \texttt{6c8e549}, \texttt{fb89ed3}, \texttt{b275a38}, \texttt{e246374},
\texttt{85c2d93}, \texttt{c1c4757}, \texttt{35648bb}, \texttt{d0a53e3}, \texttt{6fffe93}, \texttt{881a116}.
qconform-study \texttt{mutation/results/}.}\par}

\hypertarget{sec:related}{%
\section{Related work}\label{sec:related}}

\textbf{Pulse-level representations and control stacks.} The closest work is at the
same layer: a graph-based pulse representation for diverse control hardware
\cite{dalvi2024pulselib} and a pulse-level intermediate representation that follows
from it \cite{alnas2025pulseir}. Those works make one program portable across
devices. We ask whether one device can realize one program. The two compose: a
portable representation must still be checked against the device it is lowered
to. The practice of
describing a device to a program runs through the OpenPulse backend
specification \cite{mckay2018openpulse}, Qiskit Pulse \cite{alexander2020qiskitpulse},
and OpenQASM 3 \cite{cross2022openqasm3}. We evaluate on QICK \cite{stefanazzi2022qick}
and on the Qblox toolchain, which has no paper and is cited as software
\cite{qblox2026scheduler, qblox2026instruments}. QubiC \cite{xu2021qubic} and Qibolab
\cite{efthymiou2024qibolab} are other open control stacks that a descriptor could
target.

\textbf{Device and capability description.} QDMI is an interface through which a
software stack queries the properties of a device \cite{wille2024qdmi}. The Munich
Quantum Software Stack adopts it \cite{burgholzer2025mqss}, and its device model is
still being extended to new platforms \cite{stade2026devicemodels}. Our descriptor is
narrower and makes a different claim. Every limit it states cites the
observation that established it. It is pinned to a toolchain version, and a
pass of the checker is stated relative to it. A device interface could serve a
descriptor of this kind.

\textbf{Differential and metamorphic testing.} Differential testing
\cite{mckeeman1998differential}, random program generation for compilers
\cite{yang2011csmith}, and equivalence modulo inputs as an oracle \cite{le2014emi} supply
the method and its vocabulary. A survey covers the field \cite{chen2020survey}. For
quantum software, QDiff applies differential testing to quantum software stacks
\cite{wang2021qdiff}, and MorphQ applies metamorphic testing to Qiskit
\cite{paltenghi2023morphq}. An empirical study catalogs bugs in quantum computing
platforms \cite{paltenghi2022bugs}, and exact inference has been proposed as a
testing oracle \cite{lee2025exact}. Those works test at the circuit level. We test
at the pulse level against the vendor compiler.

\textbf{Companion work.} A rubric for control-plane openness measures how much of
the control surface of a device is documented and open. It has been applied to
thirteen vendor stacks \cite{malarchick2026openness}. This paper responds to the gap
that the rubric measures.

\hypertarget{sec:artifact}{%
\section{Artifact}\label{sec:artifact}}

The tool is one repository under Apache-2.0,
\url{https://github.com/rylanmalarchick/qconform}. It holds the checker, the
descriptors, the survey catalog, the corpus generator, the differential
harness, and the result rows of Section~\ref{sec:results}.
The release that this paper cites is tagged \texttt{v0.1.0}. \texttt{make\ check} builds the
checker and runs the unit tests, the golden reports, and the tripwires.
\texttt{make\ differential} regenerates the QICK result rows from the pinned toolchain, and \texttt{make\ differential-qblox}
regenerates the Qblox rows. Neither needs hardware.

The fault-injection, version, and corpus-seed studies are in a second
repository, \url{https://github.com/rylanmalarchick/qconform-study}. It pins the
tool as a submodule at the commit the results were made with. Each study runs with one
command, and each is seeded and deterministic.

\hypertarget{sec:conclusion}{%
\section{Conclusion}\label{sec:conclusion}}

qconform checks a pulse program against a versioned capability descriptor
before the program runs. Every limit in a descriptor cites the toolchain
observation behind it. The checker uses exact arithmetic, reports silent
repairs as repairs, and names the classes it did not check. A pass means that
the pinned toolchain compiles the program.

Against the QICK and Qblox toolchains, over 1263 program runs on four
configurations, the checker accepted no program that a toolchain refuses. Older
toolchain releases refuse programs that the pinned descriptors pass, on both
vendors. The corpus, the descriptors, and the result rows are the artifact, and
any implementation of the specification can be run against them.

\hypertarget{sec:ai-disclosure}{%
\section*{AI Disclosure}\label{sec:ai-disclosure}}
\addcontentsline{toc}{section}{AI Disclosure}

AI-assisted tools (Claude, Anthropic) were used for code development, running
the differential evaluation, and manuscript preparation. The research
direction, design decisions, and interpretation are the author's, who takes
full intellectual responsibility for all content.

\bibliographystyle{IEEEtran}
\bibliography{refs}

\end{document}